\documentclass[letters,fleqn,usenatbib]{mnras}

\usepackage[T1]{fontenc}

\DeclareRobustCommand{\VAN}[3]{#2}
\let\VANthebibliography\thebibliography
\def\thebibliography{\DeclareRobustCommand{\VAN}[3]{##3}\VANthebibliography}

\usepackage{graphicx}	
\usepackage{amsmath}	
\usepackage{amssymb}	

\title[Limits to Chandrasekhar-mass WD SNe Ia]
{
Chandrasekhar-mass white dwarfs are the progenitors of a small fraction of Type Ia supernovae according to nucleosythesis constraints
}

\author[E. Bravo et al.]{Eduardo Bravo$^{1}$,\thanks{E-mail: eduardo.bravo@upc.edu}
Luciano Piersanti$^{2,3}$,
St\'ephane Blondin$^{4}$,
Inma Dom\'\i nguez$^{5}$,
\newauthor Oscar Straniero$^{2,6}$,
and Sergio Cristallo$^{2,3}$
\\
$^{1}$E.T.S. Arquitectura del Vall\`es, Universitat Polit\`ecnica de Catalunya, Carrer Pere Serra 1-15, 08173 Sant Cugat del Vall\`es, Spain\\
$^{2}$INAF-Osservatorio Astronomico d'Abruzzo, via Mentore Maggini, snc, I-64100, Teramo, Italy\\
$^{3}$INFN-Sezione di Perugia, via Pascoli, Perugia, Italy\\
$^{4}$Aix Marseille Univ, CNRS, CNES, LAM, Marseille, France\\
$^{5}$Departamento de F\'\i sica Te\'orica y del Cosmos, Universidad de Granada, E-18071 Granada, Spain\\
$^{6}$INFN, Sezione di Roma, Piazzale Aldo Moro 2, 00185 Roma, Italy
}

\date{Accepted XXX. Received YYY; in original form ZZZ}

\pubyear{2022}

\begin{document}
\label{firstpage}
\pagerange{\pageref{firstpage}--\pageref{lastpage}}
\maketitle

\newcommand{\nics}{$^{56}$Ni}
\newcommand{\snia}{SNe Ia} 
\newcommand{\kms}{\,km\,s$^{-1}$} 
\newcommand{\gcc}{\,g\,cm$^{-3}$} 
\newcommand{\fecs}{$^{56}$Fe} 
\newcommand{\crcc}{$^{54}$Cr} 
\newcommand{\tict}{$^{50}$Ti} 
\newcommand{\roc}{$\rho_\mathrm{c}$} 
\newcommand{\ebg}[1]{{\color{red}{\textbf\small{#1}}}} 
\defcitealias{2022pie}{Paper~I}
\defcitealias{2013blo}{BDHK13}
\newcommand{\rev}[1]{{\color{red}{\textbf\small{#1}}}}

\begin{abstract}
The precise progenitor system of type Ia supernovae (\snia), whether it is a white dwarf  (WD) close to the Chandrasekhar limit or substantially less massive, has been a matter of debate for decades. Recent research by our group on the accretion and simmering phases preceding the explosion of a massive WD has shown that the central density at thermal runaway lies in the range $3.6 - 6.3\times10^9$\gcc\, for reasonable choices of accretion rate onto the WD and progenitor metallicity. In this work, we have computed one-dimensional simulations of the explosion of such WDs, with special emphasis on the chemical composition of the ejecta, which in all cases is extremely rich in neutronized isotopes of chromium (\crcc) and titanium (\tict). We show that, in order to reconcile such a nucleosynthesis with the isotopic abundances of the Solar System, Chandrasekhar-mass white dwarfs can account for at most 26\% of normal-luminosity \snia, or at most 20\% of all \snia. 
\end{abstract}




\section{Introduction}

Although there is consensus that Type Ia supernovae (\snia) are the result of the thermonuclear explosion of carbon-oxygen white dwarfs (WDs), the details of the evolutive path towards the explosion are still a subject of intense research: in particular, whether only a single WD takes part in the explosion \citep[single degenerate, or SD, channel,][]{1973whe} or two \citep[double degenerate, or DD, channel,][]{1984ibe,1984web}, and its implications for the evolution of \snia\, over cosmic time. Insight on the pre-eminence of one channel over the other has been provided by, e.g., the non-detection of radio and X-ray emission expected from the interaction of \snia\, with a circumstellar medium, the absence of bloated companions that should survive thousands of years after the explosion, the search for signatures of hydrogen ablated from the companion star, the time dependence of the rate of \snia\, explosions after a burst of star formation, or the search for a surviving binary companion close to \snia\, remnants. A review of the theme can be found in \citet{2014mao}.

A related but quite different issue, more directly related to the properties of the supernova, is the structure of the WD at the time of explosion: whether the exploding object is a massive WD close to its structural instability limit, i.e. a Chandrasekhar-mass WD progenitor (Ch-m WD), or its mass is smaller than $\sim1.2$~M$_{\sun}$, i.e. a sub-Chandrasekhar mass WD progenitor (subCh-m WD). Ch-m WDs can explode through two different mechanisms: a pure deflagration \citep{1982nomb} or a delayed detonation \citep{1991kho}; subCh-m WDs should detonate, and require an external trigger \citep[detonation of an accreted He-shell or a collision/merger event,][]{1994woob,2009ros,2012pak}.

The quest to determine which of Ch-m or subCh-m WDs are the dominant progenitor of \snia\, explosions has given mixed results. On the one hand, the Ch-m WD scenario is disfavoured by the failure to detect radiation associated with matter accretion onto a WD \citep{2010gil}, and by the small fraction of hydrogen ionization in \snia\, remnants \citep[SN 1572, SN 1006, and 0509-67.5,][]{2017woo,2018woo} and nebulae \citep{2019kuu}. \citet{2020flo} measure the ratio of \ion{Ni}{II} to \ion{Fe}{II} in 58 \snia\, and find that Ch-m WD explosions can only account for 11\% of their data \citep[see also][]{2022blo}, and similar conclusions are reached by \citet{2020eit} based on the evolution of the galactic abundance of manganese. \citet{2022sou} confront predictions of nebular optical line fluxes, derived from models of accreting WDs burning hydrogen or helium on their surfaces, to observations of nebulae close to supersoft X-ray sources and \snia\, remnants, and they are able to discriminate the models, leaving open just two possible accretion scenarios: direct C+O accretion onto a WD (slow merger of a DD leading to a Ch-m WD), and SD with accretion of He onto a subCh-m WD. 

On the other hand, the subCh-m WD scenario is subdominant according to analyses of the abundances in the intra-cluster medium \citep{2016mer}, and \citet{2014scab,2019sca} estimate about the same number of Ch-m and subCh-m WDs from their analysis of the light curves of several hundred \snia. 
Moreover, detailed studies of the evolution of AM CVn systems with either a He star or a He WD accreting onto a C-O WD show no clear path towards a thermonuclear explosion that might give birth to \snia\, \citep{2015bro,2015pie,2019pie}.
Here, we present new evidence of a small contribution of Ch-m WD to \snia, on the basis of pre-supernova evolution calculations followed by explosion models. 

Until now, the lack of detailed Ch-m WD evolutionary models up to the time of thermal runaway has left the ignition density largely uncertain. Simulations of the explosion and chemical evolution calculations have assumed a given central density at thermal runaway (\roc) or they have explored the effects of different densities, often exploiting the freedom to mix results from different \roc. In \citet[][hereafter Paper I]{2022pie}, we revisited the pre-explosive evolution of accreting WDs up to the onset of thermal runaway including the best available treatments of the physical processes believed to play a relevant role \citepalias[a list of prior works on the simmering phase can be found in][]{2022pie}. A key result of the new simmering calculations is that \textit{\roc\, is very high in all exploding Ch-m WDs: $\rho_\mathrm{c}>3.6\times10^9$}~\gcc. 

In this Letter, we follow the evolution through the explosive phase of the runaway WDs computed in \citetalias{2022pie}. In the following sections, we describe our methods and results, and quantify the limits posed by the nucleosynthesis to the contribution of Ch-m WDs to \snia. Our results imply that the Ch-m WD scenario cannot have been dominant among the \snia\, that contributed to the chemical enrichment of the Solar System neighborhood. Finally, we discuss the limits of our approach and give our conclusions. 

\section{Massive WD explosions}

In \citetalias{2022pie}, we show that accounting for the URCA processes associated with the most abundant species is crucial to accurately determine the structure of Ch-m WDs on the verge of explosion. In particular, the molecular weight gradient at the URCA shells confines convection to a small volume near the center,  in sharp contrast to previous belief that convection encompasses up to $\sim1$~M$_{\sun}$ of the WD. In \autoref{tab1}, we give the main properties of the computed Ch-m WDs at thermal runaway \citepalias[we adopt the model name convention of][]{2022pie}: the progenitor metallicity, $Z_\mathrm{ini}$, the accretion rate onto the exploding WD, $\dot{M}_\mathrm{acc}$, and the density of the hottest shell at thermal runaway, $\rho_{T\mathrm{max}}$, ranging from a minimum of $3.6\times10^9$~\gcc\, up to more than $6\times10^9$~\gcc. The minimum $\rho_{T\mathrm{max}}$ is obtained for high accretion rates, $\dot{M}_\mathrm{acc}\ga5\times10^{-7}~\mathrm{M}_{\sun}~\mathrm{yr}^{-1}$. 

\begin{table}
	\centering
	\caption{Model WDs at thermal runaway and explosion results.}\label{tab1}
	\begin{tabular}{lcrcrr}
		\hline
                Model & $Z_\mathrm{ini}$ & \multicolumn{1}{c}{$\dot{M}_\mathrm{acc}$} & $\rho_{T\mathrm{max}}$ & \multicolumn{1}{c}{$E_\mathrm{kin}$} & \multicolumn{1}{c}{$M_{56}$}
                \\
                 & ($10^{-3}$) & \multicolumn{1}{c}{$(\mathrm{M}_{\sun}~\mathrm{yr}^{-1})$}  & $(\mathrm{g}~\mathrm{cm}^{-3})$ &  \multicolumn{1}{c}{(foes)} & \multicolumn{1}{c}{(M$_{\sun}$)} 
                 \\
                \hline
Z14  & 0.245 & $10^{-7}$        & $5.02\times10^9$ & 1.318 & 0.656  \\         
Z63  & 6.000 & $10^{-7}$        & $5.28\times10^9$ & 1.300 & 0.642  \\        
Z12  & 13.80 & $10^{-7}$        & $5.41\times10^9$ & 1.294 & 0.619  \\
Z22  & 20.00 & $10^{-7}$        & $5.56\times10^9$ & 1.281 & 0.595  \\        
Z42  & 40.00 & $10^{-7}$        & $6.34\times10^9$ & 1.263 & 0.553  \\          
R2m7 & 13.80 & $2\times10^{-7}$ & $5.03\times10^9$ & 1.306 & 0.615  \\      
R6m7 & 13.80 & $6\times10^{-7}$ & $3.67\times10^9$ & 1.325 & 0.640  \\ 
R9m7 & 13.80 & $9\times10^{-7}$ & $3.69\times10^9$ & 1.309 & 0.615  \\         
                \hline
	\end{tabular}
	\begin{flushleft}
    Note: 1 foe=$10^{51}$~ergs\hfill
   \end{flushleft}
\end{table}

As it is well known, the outcome of the explosion of a massive WD is not uniquely determined by the initial conditions, at least not in a way that we understand well enough. We have selected delayed detonation (DDT) as the explosion mechanism, characterized by two parameters: a constant flame velocity (in terms of the sound speed) during the deflagration epoch, $v_\mathrm{def}/v_\mathrm{sound}=0.03$, and a transition density from deflagration to detonation, $\rho_\mathrm{DDT}=2.4\times10^7$~g~cm$^{-3}$. This setup has proven in past simulations to provide spectra and light curves compatible with the observed properties of normal \snia\,\citep[e.g.][although their \roc\, is lower than ours]{2013blo}. The one-dimensional hydrodynamic code used in the present calculations is described in \citet{2019bra}.

\autoref{tab1} shows the final kinetic energy, $E_\mathrm{kin}$, and the ejected mass of $^{56}$Ni, $M_{56}$.
The latter lies in the range $0.60\pm0.05$~M$_{\sun}$, close to the value obtained by \citet{2013blo} for their model DDC10, which is a close match to SN 2005cf, considered  a ``golden standard'' for typical \snia. The kinetic energy is almost the same in all models in \autoref{tab1}, but substantially smaller than the kinetic energy of model DDC10, likely due to the higher binding energy of the initial WD in the present models (DDC10 starts from a \roc\, of $2.6\times10^9$~g~cm$^{-3}$). The ejected (final, after radioactive decays) iron mass of the models in \autoref{tab1} is in the range  $0.75\pm0.01$~M$_{\sun}$, and the ratio of chromium mass to iron mass
is on the order of 0.03 in all models, about twice the Solar System ratio. However, these two elements are quite differently distributed through the ejecta: while chromium is mainly produced during the initial deflagration and concentrated in the most internal portion of the ejecta, iron is more or less evenly distributed along the deflagrated and detonated zones. Other iron-peak elements (titanium, manganese, and nickel) are underproduced or overproduced with respect to iron depending on the progenitor metallicity. 
Notably, different WD accretion rates produce almost identical chemical profiles, similar to model DDC10, and are expected to lead to indistinguishable optical outputs. 

\begin{figure}
	\includegraphics[width=\columnwidth]{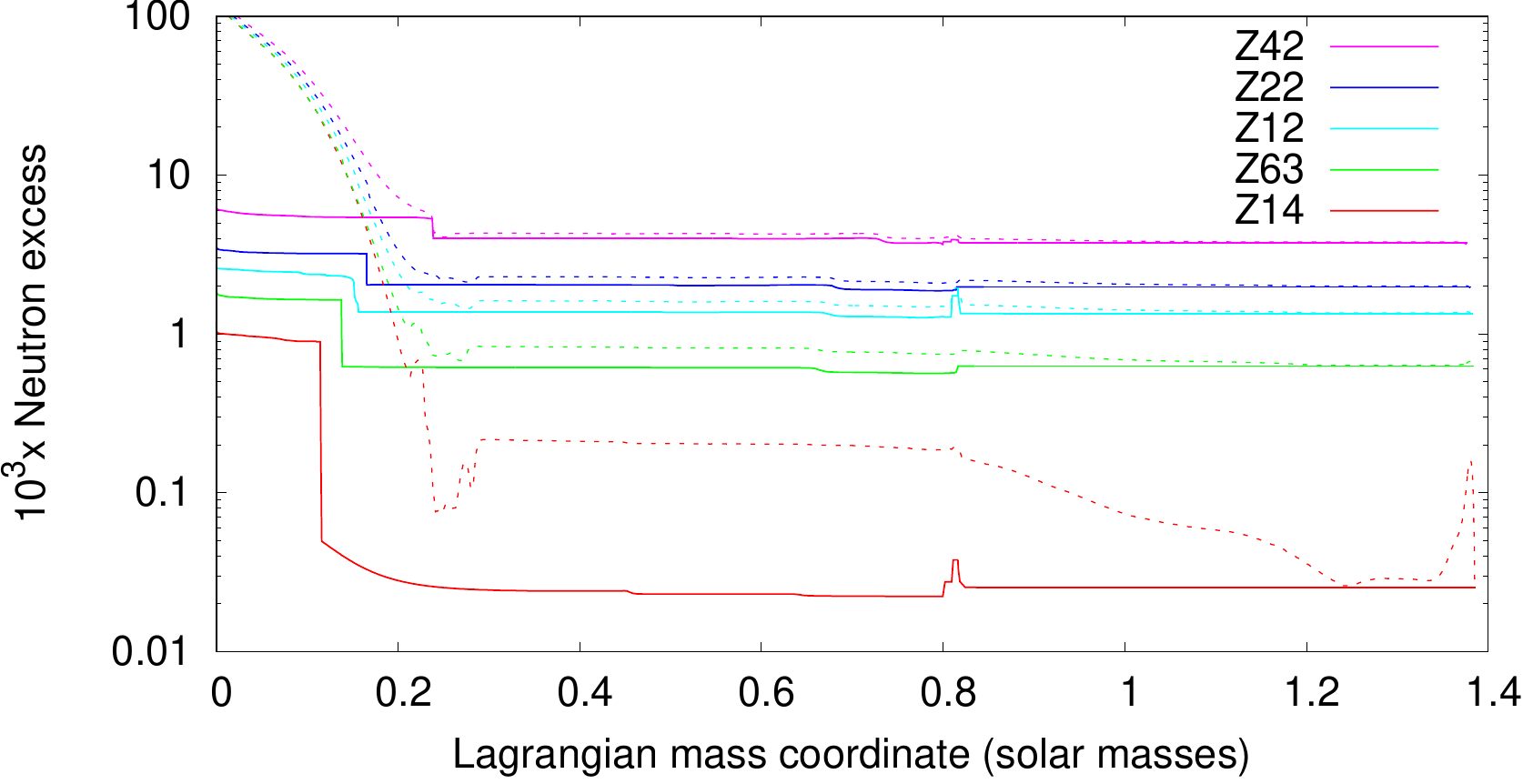}
    \caption{Profile of neutron excess, $\eta=\Sigma (N_i - Z_i)\cdot Y_i$ (where $Y_i$ is the molar fraction of species $i$) at the end of the simmering phase (solid lines) and 2,000 s after thermal runaway (dotted lines) for the models with different progenitor metallicity (Z42, Z22, Z12, Z63, and Z14, from top to bottom). The high final neutron excess in model Z14 beyond $M\sim0.25$~M$_\odot$ is due to the decay of short-lived radioactive isotopes during the early post-explosion phase. 
    }
    \label{fig-eta}
\end{figure}

The profile of the neutron excess, $\eta$, 2,000 s after thermal runaway is shown in Fig.~\ref{fig-eta}, together with the neutron excess at the end of the simmering phase. Two quite different regions can be identified in all the $\eta$ profiles. Below a mass coordinate of $\sim0.2$~M$_\odot$, the neutron excess is quite large and almost independent of the details of the initial WD structure. This behaviour is a result of the high electron capture rate in NSE at the high densities that characterize all models during the initial deflagrative phase. There, the electron capture rate is so high that all memory of the initial neutron excess is erased. Above the aforementioned mass coordinate, matter is burnt at a density small enough that there are negligible amounts of electron captures during the explosion and, consequently, the neutron excess reflects the initial metallicity of the progenitor WD. 

\begin{figure}
  \includegraphics[width=\columnwidth]{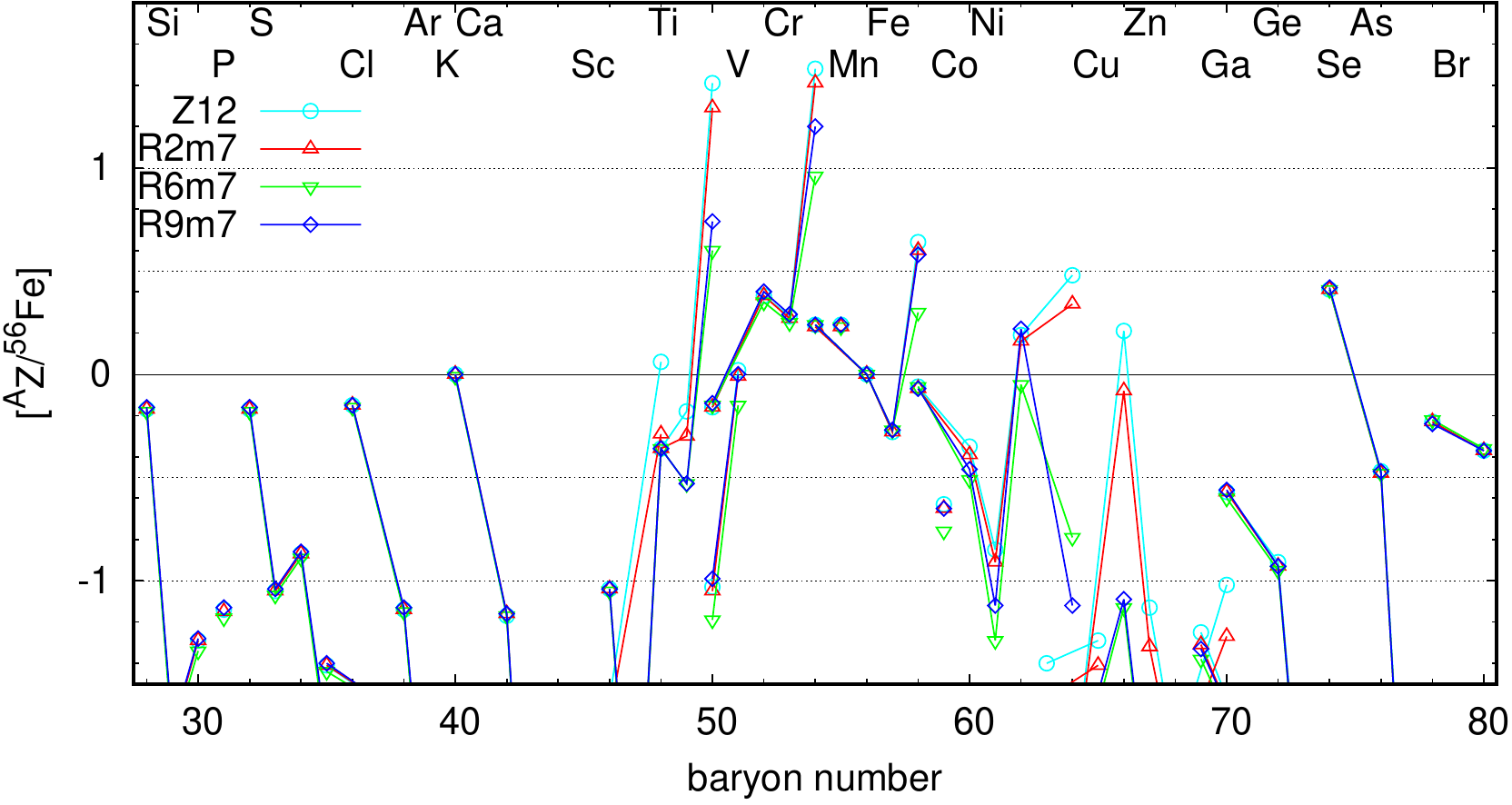}
  
  \vspace{0.2truecm}  
  
  \includegraphics[width=\columnwidth]{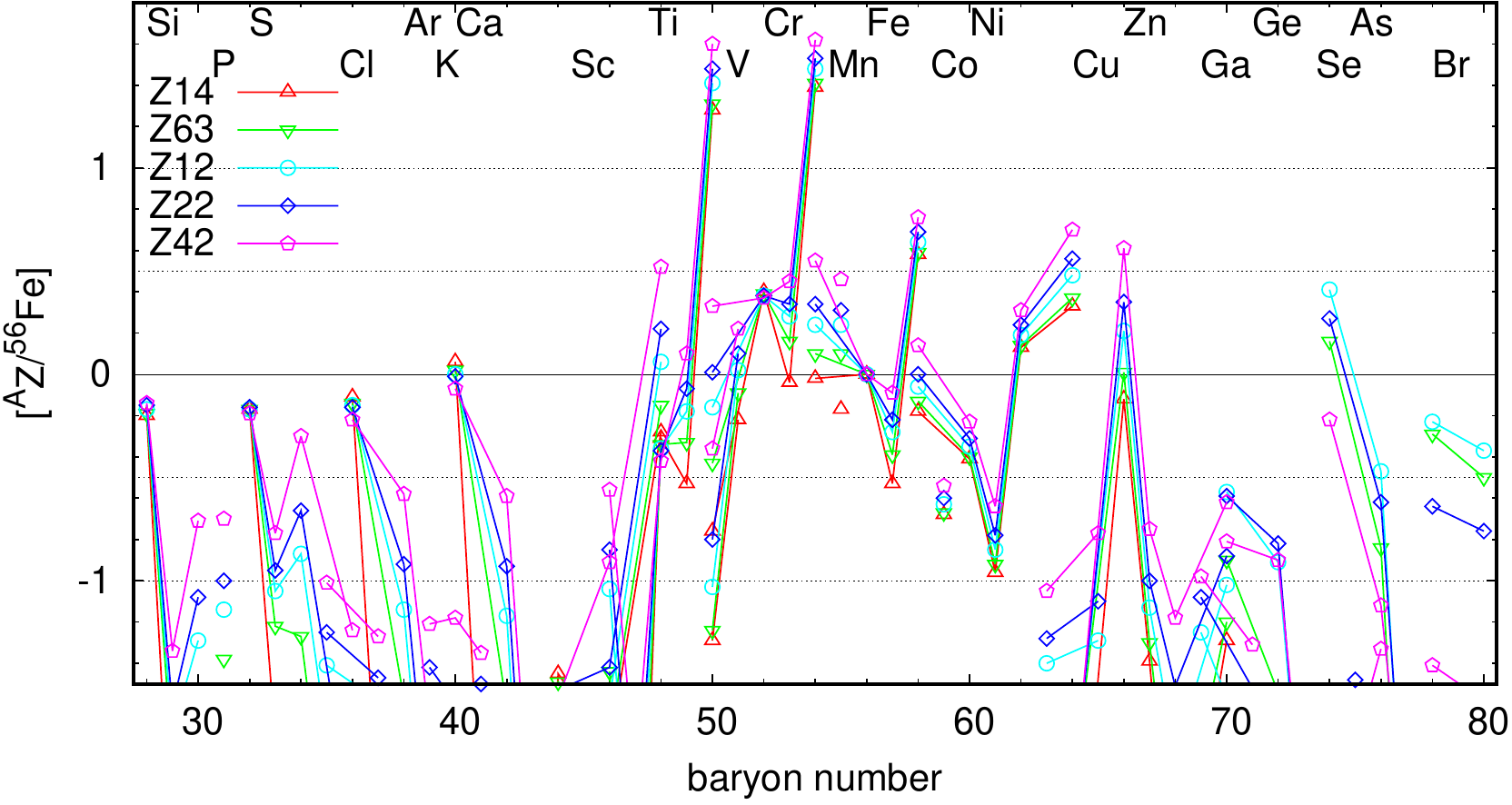}
\caption{Final composition after radioactive decays for the models with different $\dot{M}_\mathrm{acc}$ (top) and $Z_\mathrm{ini}$ (bottom): 
$
\left[^{A}Z/^{56}\mathrm{Fe}\right] =$ $\mathrm{log}_{10}\left[m\left(^{A}Z\right)/m\left(^{56}\mathrm{Fe}\right)\right]-\mathrm{log}_{10}\left[m\left(^{A}Z\right)/m\left(^{56}\mathrm{Fe}\right)\right]_\odot.
$
}
    \label{fig-compo}
\end{figure}

The chemical composition of the ejecta is shown in \autoref{fig-compo}. The first thing to note is that the composition is almost completely independent of the accretion rate onto the WD. There are, however, a few neutron-rich species that are more sensitive to the different central densities derived from the simmering evolution under different accretion rates. 
In all models, the most overproduced species (with respect to the reference, $^{56}$Fe and the Solar System composition) are $^{54}$Cr and $^{50}$Ti, with production factors above nine, even for small accretion rates. Among all the different progenitor metallicities, $Z=2.45\times10^{-4}$ produces the least amount of $^{54}$Cr, but its production factor is still on the order of 25. All the \crcc\, is produced in the region with $\eta\simeq0.1$, close to the center (see \autoref{fig-eta}). Thus, our simmering+explosion calculations show that \textit{all \snia\, coming from Ch-m WDs are characterized by a large overproduction of \crcc\, and \tict}. 
Indeed, high Cr/Fe ratios have been found in the \snia\, remnant 3C 397 \citep{2015yam,2021ohs}, and measurements of the isotopic composition of some meteoritic grains show correlated enrichments of \crcc\, and \tict\, \citep{2018nit}, in agreement with our nucleosynthetic results. 

\section{Contribution of C\lowercase{h}-\lowercase{m} WD\lowercase{s} to SN\lowercase{e} I\lowercase{a}}

The overproductions of $^{54}$Cr and $^{50}$Ti in our models limit the fraction of SNe Ia that can result from Ch-m WD explosions. Even in our most favourable case, model R6m7, and assuming that 
Ch-m 
SNe Ia 
are the only source of 
$^{54}$Cr,
only a few explosions of these massive WDs would be allowed to pollute the solar neighborhood in order to reproduce the Solar System abundances. 

In this section, we quantify upper limits to the {\sl fraction of normal SNe Ia} that can be produced by the thermonuclear explosion of {\sl Chandrasekhar-mass WDs}, which we represent with $X^\mathrm{nor}_\mathrm{Ch}$. First, we define:
$M_\mathrm{Fe}$, the total mass of Fe in the solar neighborhood;
$M_{54}$, the total mass of $^{54}$Cr in the solar neighborhood;
$f_\mathrm{Fe,Ia}$, the fraction of the Fe mass due to SNe Ia;
$X_\mathrm{sl}$, the fraction by number of subluminous SNe Ia over the total number of SNe Ia;
$m^\mathrm{nor}_\mathrm{Fe}$, the mean mass of Fe ejected in a normal SNe Ia, for both Ch-m and SubCh-m progenitors;
$m^\mathrm{sl}_\mathrm{Fe}$, the mass of Fe ejected in a subluminous SNe Ia;
$m^\mathrm{Ch}_{54}$, the mass of $^{54}$Cr ejected in a Ch-m WD SNe Ia; and
$N_\mathrm{Ia}$, the total number of SNe Ia events in the solar neighborhood.
To start, we assume that Ch-m WDs do not contribute to sub-luminous SNe Ia, and that neither core collapse SNe nor SNe Ia from subCh-m WDs synthesize any $^{54}$Cr (in order to maximize the upper limits to $X^\mathrm{nor}_\mathrm{Ch}$). Then,
\begin{equation}
 M_\mathrm{Fe} = N_\mathrm{Ia}\left[X_\mathrm{sl} m^\mathrm{sl}_\mathrm{Fe} + \left(1-X_\mathrm{sl}\right) m^\mathrm{nor}_\mathrm{Fe}\right]/f_\mathrm{Fe,Ia}\,,
\end{equation}
\noindent and
\begin{equation}
 M_{54} = N_\mathrm{Ia} X^\mathrm{nor}_\mathrm{Ch} \left(1-X_\mathrm{sl}\right) m^\mathrm{Ch}_{54}\,,
\end{equation}
\noindent Next, we constrain the ratio of the total masses of Fe and \crcc\, to match with the correspondent Solar System ratio: $M_{54}/M_\mathrm{Fe}=\left(M_{54}/M_\mathrm{Fe}\right)_{\sun}$, and we assign reasonable values to some of the above parameters: $m^\mathrm{nor}_\mathrm{Fe}=0.75$~M$_\odot$, $m^\mathrm{sl}_\mathrm{Fe}=0.25$~M$_\odot$, $X_\mathrm{sl}=0.20$ \citep{2011liw2}. 
Normal SNe Ia are expected to eject an iron mass in the range $\sim0.4-0.95$~M$_\odot$, 
with a mean close to the value we adopt for $m^\mathrm{nor}_\mathrm{Fe}$.
Although the parameter $f_\mathrm{Fe,Ia}$ is not tightly constrained, we adopt here the range from 50\% to 67\%, based on chemical evolution arguments \citep{1995tim,2018pra}. Another way to determine $f_\mathrm{Fe,Ia}$ is through the rates of different types of SNe, around one \snia\, every 3-4 core-collapse events \citep{2011liw2,2018pra,2015kub}, combined with their iron yields, about ten times larger in \snia\, than in core-collapse SNe, with the result that the contribution of \snia\, to the iron mass could be as much as 75\%. Therefore, our adopted range for $f_\mathrm{Fe,Ia}$ is quite conservative. Finally,
\begin{equation}
 X^\mathrm{nor}_\mathrm{Ch} = \frac{1 - X_\mathrm{sl} \left(1 - m^\mathrm{sl}_\mathrm{Fe}/m^\mathrm{nor}_\mathrm{Fe}\right)}{\left(1-X_\mathrm{sl}\right) f_\mathrm{Fe,Ia}\cdot\mathrm{pf}(^{54}\mathrm{Cr})}\,.\label{eq3}
\end{equation}
\noindent where $\mathrm{pf}(^{54}\mathrm{Cr})=\frac{m^\mathrm{Ch}_{54}/m^\mathrm{nor}_\mathrm{Fe}}{\left(M_{54}/M_\mathrm{Fe}\right)_\odot}$ is the production factor of $^{54}$Cr in normal-luminosity Ch-m WDs. We remark that, in order to maintain compatibility with the constraints on the contribution of SNe Ia to the chemical enrichment of the ISM, we define the production factor of \crcc\, with respect to element iron instead of $^{56}$Fe, as is common practice. In the present models, $\mathrm{pf}(^{54}\mathrm{Cr})$ goes from 8.4, for model R6m7, to 33.8, for model Z42. Applying \autoref{eq3}, the predicted fraction of Ch-m WDs among normal \snia\, goes from 5\% and up to a maximum of 26\%.

\begin{figure}
	\includegraphics[width=\columnwidth]{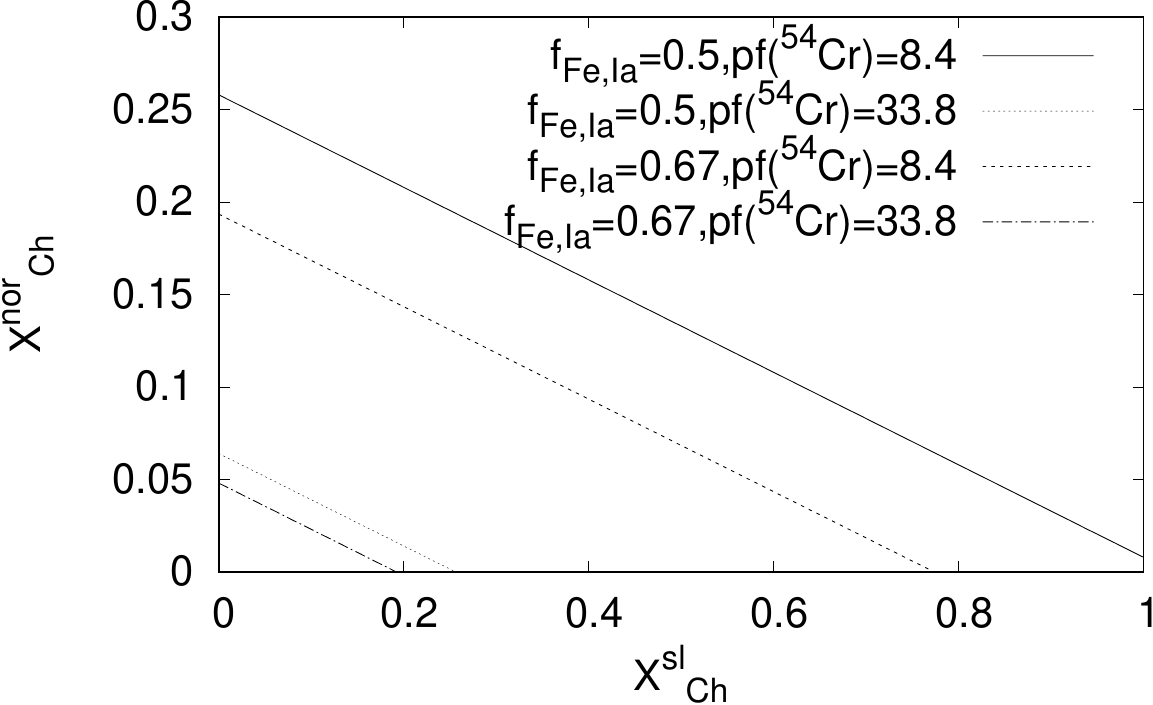}
    \caption{The fraction of normal \snia\, originating from Chandrasekhar-mass WDs, $X^\mathrm{nor}_\mathrm{Ch}$, as a function of the fraction of sub-luminous \snia\, that result from the explosion of Chandrasekhar-mass WDs, $X^\mathrm{sl}_\mathrm{Ch}$, for four combination of parameters $f_\mathrm{Fe,Ia}$ and $\mathrm{pf}($\crcc$)$.}
    \label{fig-lim}
\end{figure}

One can easily relax the assumption that the fraction of Ch-m WDs among sub-luminous \snia\,, $X^\mathrm{sl}_\mathrm{Ch}$, is different from zero. Since \crcc\, is synthesized during the first second of the deflagration phase, which is common to all the explosion mechanisms of Ch-m WDs currently considered \citep[e.g.][]{1997woo}, we assume that the yield of \crcc\, in sub-luminous \snia\, from Ch-m WDs is the same as in their normal luminosity counterparts, $m^\mathrm{Ch}_{54}$. Then, the production factor of \crcc\, in sub-luminous \snia\, from Ch-m WDs is three times larger than in normal \snia\, from the same mass range. In this case, \autoref{eq3} is replaced by a monotonouly-decreasing linear relationship between $X^\mathrm{nor}_\mathrm{Ch}$ and $X^\mathrm{sl}_\mathrm{Ch}$, shown in \autoref{fig-lim}. In the most optimistic case, $X^\mathrm{nor}_\mathrm{Ch}$ decreases from a maximum of 26\%, for $X^\mathrm{sl}_\mathrm{Ch}=0$, to a negligible value when $X^\mathrm{sl}_\mathrm{Ch}=1$, i.e. with the current uncertainties in the production factor of $^{54}$Cr and in $f_\mathrm{Fe,Ia}$, Ch-m WDs could account for all sub-luminous \snia\,, although at the cost of not contributing to normal \snia. Whatever the value of $X^\mathrm{sl}_\mathrm{Ch}$, the fraction of all \snia\, due to Ch-m WD explosions is limited to less than 20\%.

\section{Discussion}


The stellar evolution calculations presented in \citetalias{2022pie} include all the physical processes relevant to the pre-explosive WD evolution. Nevertheless, there are a few aspects that cannot or still have not been incorporated in the one-dimensional hydrostatic code, in particular rotation and magnetism, whose potential role in the simmering phase is hard to predict. Another improvement would be the modelling of the last minutes of the pre-explosive phase in three dimensions \citep[e.g.][]{2012non}, but the complexity of such a task makes it difficult to incorporate the level of physical detail that we have found relevant to the simmering phase. 

The explosive nucleosynthesis might be sensitive to both nuclear reaction uncertainties and to multidimensional hydrodynamical processes. However, the yield of \crcc\, is almost insensitive to nuclear reaction and electron capture rates \citep{2012bra,2019brab}. On the other hand, a strong buoyancy of hot ash bubbles might reduce the time that incinerated matter stay at high density and decrease the amount of electron captures, hence decreasing the yield of neutron-rich nuclei such as \crcc\, and \tict. However, the high electron capture rates at the central densities of our models translate into an increase in density of the hot bubbles in order to keep pressure constant during the almost isobaric deflagration phase, which, in turn, impacts their buoyancy. In order to estimate the last effect, we have followed the methods in \citet[][Appendix A]{2015fis} and \citet[][\S 3]{2019brab}. We find that, at densities above $5.5\times10^9$~\gcc, the bubbles are not able to float to the surface of the WD but, instead, they turn towards the center after a time $\sim0.9$~s. At slightly lower densities, $4\times10^9\la\rho\la5.5\times10^9$~\gcc, the rising of the bubbles is delayed nearly one second as compared with their evolution without electron captures. Such a delay is long enough to allow the synthesis of a large mass of \crcc\, and \tict. 

Indeed, there are a few multidimensional simulations of \snia\, starting from high-\roc\, WDs and based on the delayed detonation mechanism, and they find a large overproduction of \crcc, in line with our results. \citet{2013sei} explode a WD with a \roc\, of $5.5\times10^9$~\gcc\, and synthesize 0.69~M$_\odot$ of $^{56}$Ni (model N100H). Their production factor of $^{54}$Cr  with respect to $^{56}$Fe is five. As could be expected, the setup of model N100H is not completely consistent with our simmering models. The geometry of the initially burnt region in N100H consists of 100 spherical ignition kernels of 10 km radius distributed through a central sphere of radius 150 km, which is beyond the maximum extension of the convective core in several of our simmering models. Besides, in N100H, the initial chemical composition of the WD is 50\%-50\% carbon and oxygen, again at odds with the carbon-abundance profiles presented in \citetalias{2022pie}. \citet{2017dav} computed a high \roc\, model in two dimensions, starting from a single igniting bubble of 100 km size located at the center of a WD with a \roc\, of $6\times10^9$~\gcc. This model overproduces $^{54}$Cr by a factor $\sim50$. Besides the different dimensionality of the calculation, \citet{2017dav} assume an initially uniform chemical composition made of 30\% carbon and 70\% oxygen, close to the values of our simmering models. \citet{2018leu} compute several two-dimensional delayed detonation models with a \roc\, of $5\times10^9$~\gcc, with \crcc\, overproductions ranging from 17 to 22. 

With respect to the gravitationally confined detonation mechanism (GCD), it is based on the fast rise of a hot incinerated bubble burnt near the center to the surface of the WD \citep{2004ple}. Since we have shown that the rising of hot ash bubbles is, at the best, strongly delayed by electron captures at high \roc, we conclude that Chandrasekhar-mass WDs are not prone to explode following the GCD paradigm.

\section{Conclusions}

Our hydrodynamic explosion models of \snia\, are extremely rich in the neutronized species \crcc\, and \tict, a result that follows directly from the high \roc\, found in previous pre-explosive simmering calculations \citepalias{2022pie}. It is important to note that these simmering calculations leave no room for Ch-m WD explosions at \roc\, below $3.5\times10^9$~\gcc. Thus, we can discard approaches in which the nucleosynthesis of high-\roc\, \snia\, can be partly compensated by considering a fraction of low-\roc\, explosions, e.g. at $\sim2\times10^9$~\gcc. 

In particular, the overproduction of \crcc\, in our \snia\, models ranges from 8 (for fast accretion rates) to 34 (for slow accretion rates and super-solar metallicity). Using these overproduction factors of \crcc\, together with reasonable guesses for the contribution of \snia\, to the total mass of iron in the solar neighborhood and for the fraction of sub-luminous \snia, we derive an upper limit of $5-26\%$ to the fraction of normal \snia\, that can originate from the explosion of Ch-m WDs. The fraction of Ch-m WD progenitors over the total number of \snia\, is found to be smaller than 20\%, even when allowing for a contribution of high-density Ch-m WDs to the sub-luminous sample of \snia.

\section*{Acknowledgements}

We thank Dr. Leung for email exchange concerning published yields. E.B. thanks Amador Alvarez, Esther Nadal and the rest of the staff of CCLAIA of ETSAV for their continued support. This publication is part of the projects I + D + I PGC2018-095317-B-C21 funded by MICIN/AEI/10.13039/501100011033 and FEDER ``A way of doing Europe`` (E.B. and I.D.), and PID2021-123110NB-I00 (ID). L.P. and O.S. acknowledge financial support from 
the INAF-mainstream project ''Type Ia Supernovae 
Parent Galaxies: Expected Results from LSST``, and 
their participation to the VANS project on ''Standard candles in astrophysics: Atomic and 
Nuclear physics in SNIa``, which was funded  by the Vanvitelli University of Caserta (Italy). This work was supported by the `Programme National de Physique Stellaire' (PNPS) of CNRS/INSU co-funded by CEA and CNES (SB).
SB acknowledges support from the ESO Scientific Visitor Programme in Garching.

\section*{Data Availability}

The data underlying this article will be shared on reasonable request to the corresponding author.



\bibliographystyle{mnras}
\bibliography{../ebg.bib} 

\begin{thebibliography}{}
\makeatletter
\relax
\def\mn@urlcharsother{\let\do\@makeother \do\$\do\&\do\#\do\^\do\_\do\%\do\~}
\def\mn@doi{\begingroup\mn@urlcharsother \@ifnextchar [ {\mn@doi@}
  {\mn@doi@[]}}
\def\mn@doi@[#1]#2{\def\@tempa{#1}\ifx\@tempa\@empty \href
  {http://dx.doi.org/#2} {doi:#2}\else \href {http://dx.doi.org/#2} {#1}\fi
  \endgroup}
\def\mn@eprint#1#2{\mn@eprint@#1:#2::\@nil}
\def\mn@eprint@arXiv#1{\href {http://arxiv.org/abs/#1} {{\tt arXiv:#1}}}
\def\mn@eprint@dblp#1{\href {http://dblp.uni-trier.de/rec/bibtex/#1.xml}
  {dblp:#1}}
\def\mn@eprint@#1:#2:#3:#4\@nil{\def\@tempa {#1}\def\@tempb {#2}\def\@tempc
  {#3}\ifx \@tempc \@empty \let \@tempc \@tempb \let \@tempb \@tempa \fi \ifx
  \@tempb \@empty \def\@tempb {arXiv}\fi \@ifundefined
  {mn@eprint@\@tempb}{\@tempb:\@tempc}{\expandafter \expandafter \csname
  mn@eprint@\@tempb\endcsname \expandafter{\@tempc}}}

\bibitem[\protect\citeauthoryear{{Blondin}, {Dessart}, {Hillier}  \&
  {Khokhlov}}{{Blondin} et~al.}{2013}]{2013blo}
{Blondin} S.,  {Dessart} L.,  {Hillier} D.~J.,   {Khokhlov} A.~M.,  2013,
  \mn@doi [\mnras] {10.1093/mnras/sts484}, \href
  {http://cdsads.u-strasbg.fr/abs/2013MNRAS.429.2127B} {429, 2127}

\bibitem[\protect\citeauthoryear{{Blondin}, {Bravo}, {Timmes}, {Dessart}  \&
  {Hillier}}{{Blondin} et~al.}{2022}]{2022blo}
{Blondin} S.,  {Bravo} E.,  {Timmes} F.~X.,  {Dessart} L.,   {Hillier} D.~J.,
  2022, \mn@doi [\aap] {10.1051/0004-6361/202142323}, \href
  {https://ui.adsabs.harvard.edu/abs/2022A&A...660A..96B} {660, A96}

\bibitem[\protect\citeauthoryear{{Bravo}}{{Bravo}}{2019}]{2019brab}
{Bravo} E.,  2019, \mn@doi [\aap] {10.1051/0004-6361/201935095}, \href
  {https://ui.adsabs.harvard.edu/abs/2019A%26A...624A.139B} {624, A139}

\bibitem[\protect\citeauthoryear{{Bravo} \& {Mart{\'i}nez-Pinedo}}{{Bravo} \&
  {Mart{\'i}nez-Pinedo}}{2012}]{2012bra}
{Bravo} E.,  {Mart{\'i}nez-Pinedo} G.,  2012, \mn@doi [\prc]
  {10.1103/PhysRevC.85.055805}, \href
  {http://cdsads.u-strasbg.fr/abs/2012PhRvC..85e5805B} {85, 055805}

\bibitem[\protect\citeauthoryear{{Bravo}, {Badenes}  \&
  {Mart{\'{\i}}nez-Rodr{\'{\i}}guez}}{{Bravo} et~al.}{2019}]{2019bra}
{Bravo} E.,  {Badenes} C.,   {Mart{\'{\i}}nez-Rodr{\'{\i}}guez} H.,  2019,
  \mn@doi [\mnras] {10.1093/mnras/sty2951}, \href
  {http://cdsads.u-strasbg.fr/abs/2019MNRAS.482.4346B} {482, 4346}

\bibitem[\protect\citeauthoryear{{Brooks}, {Bildsten}, {Marchant}  \&
  {Paxton}}{{Brooks} et~al.}{2015}]{2015bro}
{Brooks} J.,  {Bildsten} L.,  {Marchant} P.,   {Paxton} B.,  2015, \mn@doi
  [\apj] {10.1088/0004-637X/807/1/74}, \href
  {https://ui.adsabs.harvard.edu/abs/2015ApJ...807...74B} {807, 74}

\bibitem[\protect\citeauthoryear{{Dave}, {Kashyap}, {Fisher}, {Timmes},
  {Townsley}  \& {Byrohl}}{{Dave} et~al.}{2017}]{2017dav}
{Dave} P.,  {Kashyap} R.,  {Fisher} R.,  {Timmes} F.,  {Townsley} D.,
  {Byrohl} C.,  2017, \mn@doi [\apj] {10.3847/1538-4357/aa7134}, \href
  {http://cdsads.u-strasbg.fr/abs/2017ApJ...841...58D} {841, 58}

\bibitem[\protect\citeauthoryear{{Eitner}, {Bergemann}, {Hansen}, {Cescutti},
  {Seitenzahl}, {Larsen}  \& {Plez}}{{Eitner} et~al.}{2020}]{2020eit}
{Eitner} P.,  {Bergemann} M.,  {Hansen} C.~J.,  {Cescutti} G.,  {Seitenzahl}
  I.~R.,  {Larsen} S.,   {Plez} B.,  2020, \mn@doi [\aap]
  {10.1051/0004-6361/201936603}, \href
  {https://ui.adsabs.harvard.edu/abs/2020A&A...635A..38E} {635, A38}

\bibitem[\protect\citeauthoryear{{Fisher} \& {Jumper}}{{Fisher} \&
  {Jumper}}{2015}]{2015fis}
{Fisher} R.,  {Jumper} K.,  2015, \mn@doi [\apj] {10.1088/0004-637X/805/2/150},
  \href {http://cdsads.u-strasbg.fr/abs/2015ApJ...805..150F} {805, 150}

\bibitem[\protect\citeauthoryear{{Fl{\"o}rs} et~al.,}{{Fl{\"o}rs}
  et~al.}{2020}]{2020flo}
{Fl{\"o}rs} A.,  et~al., 2020, \mn@doi [\mnras] {10.1093/mnras/stz3013}, \href
  {https://ui.adsabs.harvard.edu/abs/2020MNRAS.491.2902F} {491, 2902}

\bibitem[\protect\citeauthoryear{{Gilfanov} \& {Bogd{\'a}n}}{{Gilfanov} \&
  {Bogd{\'a}n}}{2010}]{2010gil}
{Gilfanov} M.,  {Bogd{\'a}n} {\'A}.,  2010, \mn@doi [\nat]
  {10.1038/nature08685}, \href
  {https://ui.adsabs.harvard.edu/abs/2010Natur.463..924G} {463, 924}

\bibitem[\protect\citeauthoryear{{Iben} \& {Tutukov}}{{Iben} \&
  {Tutukov}}{1984}]{1984ibe}
{Iben} Jr. I.,  {Tutukov} A.~V.,  1984, \mn@doi [\apjs] {10.1086/190932}, \href
  {http://cdsads.u-strasbg.fr/abs/1984ApJS...54..335I} {54, 335}

\bibitem[\protect\citeauthoryear{{Khokhlov}}{{Khokhlov}}{1991}]{1991kho}
{Khokhlov} A.~M.,  1991, \aap, \href
  {http://cdsads.u-strasbg.fr/abs/1991A\%26A...245..114K} {245, 114}

\bibitem[\protect\citeauthoryear{{Kubryk}, {Prantzos}  \&
  {Athanassoula}}{{Kubryk} et~al.}{2015}]{2015kub}
{Kubryk} M.,  {Prantzos} N.,   {Athanassoula} E.,  2015, \mn@doi [\aap]
  {10.1051/0004-6361/201424171}, \href
  {https://ui.adsabs.harvard.edu/abs/2015A&A...580A.126K} {580, A126}

\bibitem[\protect\citeauthoryear{{Kuuttila}, {Gilfanov}, {Seitenzahl}, {Woods}
  \& {Vogt}}{{Kuuttila} et~al.}{2019}]{2019kuu}
{Kuuttila} J.,  {Gilfanov} M.,  {Seitenzahl} I.~R.,  {Woods} T.~E.,   {Vogt}
  F.~P.~A.,  2019, \mn@doi [\mnras] {10.1093/mnras/stz065}, \href
  {https://ui.adsabs.harvard.edu/abs/2019MNRAS.484.1317K} {484, 1317}

\bibitem[\protect\citeauthoryear{{Leung} \& {Nomoto}}{{Leung} \&
  {Nomoto}}{2018}]{2018leu}
{Leung} S.-C.,  {Nomoto} K.,  2018, \mn@doi [\apj] {10.3847/1538-4357/aac2df},
  \href {https://ui.adsabs.harvard.edu/abs/2018ApJ...861..143L} {861, 143}

\bibitem[\protect\citeauthoryear{Li et~al.,}{Li et~al.}{2011}]{2011liw2}
Li W.,  et~al., 2011, \mn@doi [Monthly Notices of the Royal Astronomical
  Society] {10.1111/j.1365-2966.2011.18160.x}, 412, 1441

\bibitem[\protect\citeauthoryear{Maoz, Mannucci  \& Nelemans}{Maoz
  et~al.}{2014}]{2014mao}
Maoz D.,  Mannucci F.,   Nelemans G.,  2014, \mn@doi [Annual Review of
  Astronomy and Astrophysics] {10.1146/annurev-astro-082812-141031}, 52, 107

\bibitem[\protect\citeauthoryear{{Mernier} et~al.,}{{Mernier}
  et~al.}{2016}]{2016mer}
{Mernier} F.,  et~al., 2016, \mn@doi [\aap] {10.1051/0004-6361/201628765},
  \href {https://ui.adsabs.harvard.edu/abs/2016A&A...595A.126M} {595, A126}

\bibitem[\protect\citeauthoryear{{Nittler}, {Alexander}, {Liu}  \&
  {Wang}}{{Nittler} et~al.}{2018}]{2018nit}
{Nittler} L.~R.,  {Alexander} C. M.~O.,  {Liu} N.,   {Wang} J.,  2018, \mn@doi
  [\apjl] {10.3847/2041-8213/aab61f}, \href
  {https://ui.adsabs.harvard.edu/abs/2018ApJ...856L..24N} {856, L24}

\bibitem[\protect\citeauthoryear{{Nomoto}}{{Nomoto}}{1982}]{1982nomb}
{Nomoto} K.,  1982, \mn@doi [\apj] {10.1086/159682}, \href
  {http://cdsads.u-strasbg.fr/abs/1982ApJ...253..798N} {253, 798}

\bibitem[\protect\citeauthoryear{{Nonaka}, {Aspden}, {Zingale}, {Almgren},
  {Bell}  \& {Woosley}}{{Nonaka} et~al.}{2012}]{2012non}
{Nonaka} A.,  {Aspden} A.~J.,  {Zingale} M.,  {Almgren} A.~S.,  {Bell} J.~B.,
  {Woosley} S.~E.,  2012, \mn@doi [\apj] {10.1088/0004-637X/745/1/73}, \href
  {https://ui.adsabs.harvard.edu/abs/2012ApJ...745...73N} {745, 73}

\bibitem[\protect\citeauthoryear{{Ohshiro} et~al.,}{{Ohshiro}
  et~al.}{2021}]{2021ohs}
{Ohshiro} Y.,  et~al., 2021, \mn@doi [\apjl] {10.3847/2041-8213/abff5b}, \href
  {https://ui.adsabs.harvard.edu/abs/2021ApJ...913L..34O} {913, L34}

\bibitem[\protect\citeauthoryear{{Pakmor}, {Kromer}, {Taubenberger}, {Sim},
  {R{\"o}pke}  \& {Hillebrandt}}{{Pakmor} et~al.}{2012}]{2012pak}
{Pakmor} R.,  {Kromer} M.,  {Taubenberger} S.,  {Sim} S.~A.,  {R{\"o}pke}
  F.~K.,   {Hillebrandt} W.,  2012, \mn@doi [\apjl]
  {10.1088/2041-8205/747/1/L10}, \href
  {https://ui.adsabs.harvard.edu/abs/2012ApJ...747L..10P} {747, L10}

\bibitem[\protect\citeauthoryear{{Piersanti}, {Yungelson}  \&
  {Tornamb{\'e}}}{{Piersanti} et~al.}{2015}]{2015pie}
{Piersanti} L.,  {Yungelson} L.~R.,   {Tornamb{\'e}} A.,  2015, \mn@doi
  [\mnras] {10.1093/mnras/stv1452}, \href
  {https://ui.adsabs.harvard.edu/abs/2015MNRAS.452.2897P} {452, 2897}

\bibitem[\protect\citeauthoryear{{Piersanti}, {Yungelson}, {Cristallo}  \&
  {Tornamb{\'e}}}{{Piersanti} et~al.}{2019}]{2019pie}
{Piersanti} L.,  {Yungelson} L.~R.,  {Cristallo} S.,   {Tornamb{\'e}} A.,
  2019, \mn@doi [\mnras] {10.1093/mnras/stz033}, \href
  {https://ui.adsabs.harvard.edu/abs/2019MNRAS.484..950P} {484, 950}

\bibitem[\protect\citeauthoryear{{Piersanti}, {Bravo}, {Straniero}, {Cristallo}
   \& {Dom{\'\i}nguez}}{{Piersanti} et~al.}{2022}]{2022pie}
{Piersanti} L.,  {Bravo} E.,  {Straniero} O.,  {Cristallo} S.,
  {Dom{\'\i}nguez} I.,  2022, \mn@doi [\apj] {10.3847/1538-4357/ac403b}, \href
  {https://ui.adsabs.harvard.edu/abs/2022ApJ...926..103P} {926, 103 (Paper I)}

\bibitem[\protect\citeauthoryear{{Plewa}, {Calder}  \& {Lamb}}{{Plewa}
  et~al.}{2004}]{2004ple}
{Plewa} T.,  {Calder} A.~C.,   {Lamb} D.~Q.,  2004, \mn@doi [\apjl]
  {10.1086/424036}, \href {http://cdsads.u-strasbg.fr/abs/2004ApJ...612L..37P}
  {612, L37}

\bibitem[\protect\citeauthoryear{{Prantzos}, {Abia}, {Limongi}, {Chieffi}  \&
  {Cristallo}}{{Prantzos} et~al.}{2018}]{2018pra}
{Prantzos} N.,  {Abia} C.,  {Limongi} M.,  {Chieffi} A.,   {Cristallo} S.,
  2018, \mn@doi [\mnras] {10.1093/mnras/sty316}, \href
  {http://cdsads.u-strasbg.fr/abs/2018MNRAS.476.3432P} {476, 3432}

\bibitem[\protect\citeauthoryear{{Rosswog}, {Kasen}, {Guillochon}  \&
  {Ramirez-Ruiz}}{{Rosswog} et~al.}{2009}]{2009ros}
{Rosswog} S.,  {Kasen} D.,  {Guillochon} J.,   {Ramirez-Ruiz} E.,  2009,
  \mn@doi [\apjl] {10.1088/0004-637X/705/2/L128}, \href
  {http://cdsads.u-strasbg.fr/abs/2009ApJ...705L.128R} {705, L128}

\bibitem[\protect\citeauthoryear{{Scalzo}, {Ruiter}  \& {Sim}}{{Scalzo}
  et~al.}{2014}]{2014scab}
{Scalzo} R.~A.,  {Ruiter} A.~J.,   {Sim} S.~A.,  2014, \mn@doi [\mnras]
  {10.1093/mnras/stu1808}, \href
  {https://ui.adsabs.harvard.edu/abs/2014MNRAS.445.2535S} {445, 2535}

\bibitem[\protect\citeauthoryear{{Scalzo} et~al.,}{{Scalzo}
  et~al.}{2019}]{2019sca}
{Scalzo} R.~A.,  et~al., 2019, \mn@doi [\mnras] {10.1093/mnras/sty3178}, \href
  {https://ui.adsabs.harvard.edu/abs/2019MNRAS.483..628S} {483, 628}

\bibitem[\protect\citeauthoryear{{Seitenzahl} et~al.,}{{Seitenzahl}
  et~al.}{2013}]{2013sei}
{Seitenzahl} I.~R.,  et~al., 2013, \mn@doi [\mnras] {10.1093/mnras/sts402},
  \href {https://ui.adsabs.harvard.edu/abs/2013MNRAS.429.1156S} {429, 1156}

\bibitem[\protect\citeauthoryear{{Souropanis}, {Chiotellis}, {Boumis},
  {Chatzikos}, {Akras}, {Piersanti}, {Ruiter}  \& {Ferland}}{{Souropanis}
  et~al.}{2022}]{2022sou}
{Souropanis} D.,  {Chiotellis} A.,  {Boumis} P.,  {Chatzikos} M.,  {Akras} S.,
  {Piersanti} L.,  {Ruiter} A.~J.,   {Ferland} G.~J.,  2022, \mn@doi [\mnras]
  {10.1093/mnras/stac890}, \href
  {https://ui.adsabs.harvard.edu/abs/2022MNRAS.513.2369S} {513, 2369}

\bibitem[\protect\citeauthoryear{{Timmes}, {Woosley}  \& {Weaver}}{{Timmes}
  et~al.}{1995}]{1995tim}
{Timmes} F.~X.,  {Woosley} S.~E.,   {Weaver} T.~A.,  1995, \mn@doi [\apjs]
  {10.1086/192172}, \href {http://cdsads.u-strasbg.fr/abs/1995ApJS...98..617T}
  {98, 617}

\bibitem[\protect\citeauthoryear{{Webbink}}{{Webbink}}{1984}]{1984web}
{Webbink} R.~F.,  1984, \mn@doi [\apj] {10.1086/161701}, \href
  {https://ui.adsabs.harvard.edu/abs/1984ApJ...277..355W} {277, 355}

\bibitem[\protect\citeauthoryear{{Whelan} \& {Iben}}{{Whelan} \&
  {Iben}}{1973}]{1973whe}
{Whelan} J.,  {Iben} Icko J.,  1973, \mn@doi [\apj] {10.1086/152565}, \href
  {https://ui.adsabs.harvard.edu/abs/1973ApJ...186.1007W} {186, 1007}

\bibitem[\protect\citeauthoryear{{Woods}, {Ghavamian}, {Badenes}  \&
  {Gilfanov}}{{Woods} et~al.}{2017}]{2017woo}
{Woods} T.~E.,  {Ghavamian} P.,  {Badenes} C.,   {Gilfanov} M.,  2017, \mn@doi
  [Nature Astronomy] {10.1038/s41550-017-0263-5}, \href
  {https://ui.adsabs.harvard.edu/abs/2017NatAs...1..800W} {1, 800}

\bibitem[\protect\citeauthoryear{{Woods}, {Ghavamian}, {Badenes}  \&
  {Gilfanov}}{{Woods} et~al.}{2018}]{2018woo}
{Woods} T.~E.,  {Ghavamian} P.,  {Badenes} C.,   {Gilfanov} M.,  2018, \mn@doi
  [\apj] {10.3847/1538-4357/aad1ee}, \href
  {https://ui.adsabs.harvard.edu/abs/2018ApJ...863..120W} {863, 120}

\bibitem[\protect\citeauthoryear{{Woosley}}{{Woosley}}{1997}]{1997woo}
{Woosley} S.~E.,  1997, \mn@doi [\apj] {10.1086/303650}, \href
  {https://ui.adsabs.harvard.edu/abs/1997ApJ...476..801W} {476, 801}

\bibitem[\protect\citeauthoryear{{Woosley} \& {Weaver}}{{Woosley} \&
  {Weaver}}{1994}]{1994woob}
{Woosley} S.~E.,  {Weaver} T.~A.,  1994, \mn@doi [\apj] {10.1086/173813}, \href
  {http://cdsads.u-strasbg.fr/abs/1994ApJ...423..371W} {423, 371}

\bibitem[\protect\citeauthoryear{{Yamaguchi} et~al.,}{{Yamaguchi}
  et~al.}{2015}]{2015yam}
{Yamaguchi} H.,  et~al., 2015, \mn@doi [\apjl] {10.1088/2041-8205/801/2/L31},
  \href {http://cdsads.u-strasbg.fr/abs/2015ApJ...801L..31Y} {801, L31}

\makeatother
\end{thebibliography}








\bsp	
\label{lastpage}
\end{document}